\documentclass[twocolumn,showpacs,amsmath,amssymb,aps,superscriptaddress,nofootinbib,longbibliography]{revtex4-1}

\usepackage{graphicx}
\usepackage{multirow}
\usepackage{amsmath}
\usepackage{amsthm}
\usepackage{soul}
\usepackage{verbatim}
\usepackage{bm}
\usepackage{outlines}
\usepackage{siunitx}
\usepackage[space]{grffile}
\usepackage[colorlinks=true, linkcolor=blue, citecolor=gray]{hyperref}
\usepackage[capitalize]{cleveref}
\usepackage{xr}
\usepackage{color}
\usepackage{tikz}
\usetikzlibrary{calc, positioning}
\usepackage{algorithm}
\usepackage{algpseudocode}

\newcommand{\PP}{\mathbb{P}}
\newcommand{\Oo}{\mathcal{O}}
\newcommand{\Pp}{\mathcal{P}}

\newcommand{\ZZ}{\mathbb{Z}}

\def\avg#1{\mathinner{\langle{#1}\rangle}}

\newtheorem{thm}{Theorem}

\begin{document}

\title{Nearly Optimal Measurement Scheduling for Partial Tomography of Quantum States}

\begin{abstract}
Many applications of quantum simulation require to prepare and then characterize quantum states by efficiently estimating $k$-body reduced density matrices ($k$-RDMs), from which observables of interest may be obtained.
For instance, the fermionic $2$-RDM contains the energy, charge density, and energy gradients of an electronic system, while the qubit $2$-RDM contains the spatial correlation functions of magnetic systems.
Naive estimation of such RDMs require repeated state preparations for each matrix element, which makes for prohibitively large computation times.
However, commuting matrix elements may be measured simultaneously, allowing for a significant cost reduction.
In this work we design schemes for such parallelization with near-optimal complexity in the system size $N$.
We first describe a scheme to sample all elements of a qubit $k$-RDM using only ${\cal O}(3^{k} \log^{k-1}\! N)$ unique measurement circuits, an exponential improvement over prior art.
We then describe a scheme for sampling all elements of the fermionic 2-RDM using only ${\cal O}(N^2)$ unique measurement circuits, each of which requires only a local ${\cal O}(N)$-depth measurement circuit.
We prove a lower bound of $\Omega(\epsilon^{-2}N^k)$ on the number of state preparations, Clifford circuits, and measurement in the computational basis required to estimate all elements of a fermionic $k$-RDM, making our scheme for sampling the fermionic 2-RDM asymptotically optimal.
We finally construct circuits to sample the expectation value of a linear combination of $\omega$ anti-commuting $2$-body fermionic operators with only ${\cal O}(\omega)$ gates on a linear array. 
This allows for sampling any linear combination of fermionic 2-RDM elements in ${\cal O}(N^4 / \omega)$ time, with a significantly lower measurement circuit complexity than prior art.
Our results improve the viability of near-term quantum simulation of molecules and strongly correlated material systems.
\end{abstract}

\date{\today}

\author{Xavier Bonet-Monroig}
\affiliation{Instituut-Lorentz, Universiteit Leiden, 2300 RA Leiden, The Netherlands}
\affiliation{QuTech, Delft University of Technology, 2600 GA Delft, The Netherlands}

\author{Ryan Babbush}
\affiliation{Google Research, Venice, CA 90291, United States}

\author{Thomas O'Brien}
\affiliation{Instituut-Lorentz, Universiteit Leiden, 2300 RA Leiden, The Netherlands}
\affiliation{Google Research, Venice, CA 90291, United States}

\maketitle

\section{Introduction}
The advent of variational methods, most notably the variational quantum eigensolver~\cite{peruzzo2014variational,mcclean2016theory}, inspires hope that useful contributions to our understanding of strongly-correlated physical and chemical systems might be achievable in pre-error corrected quantum devices~\cite{omalley2016scalable}.
Following this initial work, much progress has gone into lowering the coherence requirements of variational methods~\cite{romero2018strategies}, calculating system properties beyond ground state energies~\cite{mcclean2017hybrid,mcardle2018vibration,obrien2019calculating}, and experimental implementation~\cite{kandala2017hardware,colless2018computation,Hempel2018,Nam2019}.
However, extracting data from an exponentially complex quantum state is a critical bottleneck for such applications.
Initial estimates for the number of measurements required to accurately approximate the energy of a variationally generated quantum state were astronomically large, with bounds for quantum chemistry applications as high as $10^{13}$ for a system of 112 spin-orbitals in minimal basis~\cite{wecker2015progress}.
Although improving these results is critical for the scalability of variational approaches, until recently, little effort has been devoted to lowering the number of measurements needed.

A common way to estimate the energy of a quantum state during a variational quantum algorithm is to perform partial tomography~\cite{mcclean2016theory} on a set of observables which comprise a $k$-body reduced density matrix ($k$-RDM)\footnote{While $k$-body qubit RDMs catalogue correlations between $k$ qubits, $k$-body fermion RDMs catalogue correlations between $k$ fermions, and thus involve $2k$ fermionic modes; e.g., the elements of the fermionic 2-RDM are the expectation values $\avg{c^\dagger_p c^\dagger_q c_r c_s}$.}~\cite{rubin2018application}.
For instance, the fermionic 2-RDM allows one to calculate such properties as energy~\cite{rubin2018application}, energy gradients~\cite{overy2014unbiased,obrien2019calculating}, and multipole moments~\cite{gidofalvi2007molecular} of electronic systems in quantum chemistry and condensed matter problems, and further enables techniques for relaxing orbitals to reduce basis error~\cite{mcclean2017hybrid,Takeshita2019}.
By contrast, the qubit 2-RDM plays a vital role in spin systems, as it contains static spin correlation functions that can be used to predict phases and phase transitions~\cite{sethna2006correlations}, and separately contains information to characterize the entanglement generated on a quantum device~\cite{cotler2019quantum}. 
Reduced density matrices thus offer a useful and tractable description of an otherwise complex quantum state.

Partial tomography to estimate a reduced density matrix may be performed by separating the observables to be tomographed into sets of mutually-commuting operators.
By virtue of their commutation, a unique measurement scheme may be found to measure all operators in a single set simultaneously.
Subsequent measurement of non-commuting operators requires re-preparation of the quantum state, so the time required to estimate a target RDM is proportional to the number of unique measurement circuits.
Minimizing this number is crucial for the scalibility of variational algorithms, as a naive approach requires ${\cal O}(N^4)$ unique measurement circuits, which is impractical.
Recent work has focused on mapping this problem to that of clique finding or colouring of a graph~\cite{verteletskyi2019measurement}, and applying approximate algorithms to these known NP-hard problems~\cite{karp1972reducibility}.
This achieves constant or empirically determined linear scaling improvements over an approach that measures each term individually~\cite{verteletskyi2019measurement,jena2019pauli,yen2019measuring,gokhale2019minimizing,izmaylov2019unitary}.
However, the commutation relations between local qubit or local fermionic operators has significant regularity not utilised in naive graph-theoretic algorithms.
Leveraging this regularity is critical to optimizing and proving bounds on the difficulty of tomography of quantum states.

\begin{table*}[t!]
\begin{tabular}{c|c|c|c|c|c|c|c|c|c}

ref. & partitioning method & circuits based on & partitions & gate count  & depth & classical cost & connect. & RDM & sym. \\
\hline\hline
\cite{mcclean2016theory} & comm. Pauli heuristic & - & ${\cal O}(N^4)$ & - & - & ${\cal O}(1)$ & - & - & -\\
\cite{kandala2017hardware}
& compatible Pauli heuristic & single rotations & ${\cal O}(N^4)$ & $N$ & 1 & ${\cal O}(1)$ & linear & yes & no\\
\cite{rubin2018application} 
& $n$-representability constraints & single rotations & ${\cal O}(N^4)$ & $N$ & 1 & ${\cal O}(1)$ & linear & no & no\\
\cite{izmaylov2018revising}
& mean-field partitioning & fast feed-forward & ${\cal O}(N^4)$ & ${\cal O}(N)$ & ${\cal O}(N)$ & ${\cal O}(N^3)$ & full & no & no\\
\cite{verteletskyi2019measurement}
& compatible Pauli clique cover & single rotations & ${\cal O}(N^4)$ & $N$ & $1$ & ${\cal O}(N^8-N^{12})$ & linear & yes & no\\
\cite{jena2019pauli}
& comm. Pauli graph coloring  & stabilizer formalism & ${\cal O}(N^3)$ & - & - & - & full & no & no\\
 \cite{izmaylov2019unitary} 
 & a-comm. Pauli clique cover & Pauli evolutions & ${\cal O}(N^3)$ & ${\cal O}(N^2 \log N)$ & - &  ${\cal O}(N^8-N^{12})$ & full & no & no\\
 \cite{yen2019measuring}
 & comm. Pauli clique cover & symplectic subspaces & ${\cal O}(N^3)$ & ${\cal O}(N^2 / \log N)$ & - & $O(N^8-N^{12})$ & full & no & no\\
  \cite{huggins2019efficient} & basis rotation grouping 
 & Givens rotations & ${\cal O}(N)$ & $N^2 / 4$ & $N / 2$ & ${\cal O}(N^4\log(N))$ & linear & no & Num.\\
\cite{gokhale2019minimizing}
& comm. Pauli clique cover & stabilizer formalism & ${\cal O}(N^3)$ &  ${\cal O}(N^2)$ & - &  ${\cal O}(N^8-N^{12})$ & full & yes & no\\
\cite{crawford2019efficient}
& comm. Pauli clique cover & stabilizer formalism & ${\cal O}(N^3)$ &  ${\cal O}(N^2)$ & - &  ${\cal O}(N^8-N^{12})$ & full & yes & no\\
\cite{zhao2019measurement}
 & a-comm. Pauli clique cover & Pauli evolutions & ${\cal O}(N^3)$ & ${\cal O}(N^{\frac{3}{2}} \log N)$ & - & ${\cal O}(N^8-N^{12})$ & full & no & no\\
 here & comm. Majorana pairs & Majorana swaps & $\Oo(N^2)$ & $N^2 / 2$ & $N$ & ${\cal O}(N^2)$ & linear & yes & Par.\\
 here & a-comm. Majoranas & Majorana rotations & ${\cal O}(N^4 / \omega) $ & $\omega$ & $\omega / 2$ & ${\cal O}(\frac{N^4}{ \omega})$ & linear & no & Par.\\
 here & 2-RDM partition bound & - & $\Omega(N^2)$ & - & - & - & - & - & - \\
 here & a-comm. clique bound & - & $\Omega(N^3)$ & - & - & - & - & - & -\\
\hline
\end{tabular}
\caption{\label{tab:comparison}A history of ideas reducing the measurements required for estimating the energy of arbitrary basis chemistry Hamiltonians with the variational quantum eigensolver.
Here $N$ represents the number of spin-orbitals in the basis, and $\omega$ is defined in the text.
We use the shorthand ``comm''. and ``a-comm.'' for commuting and anti-commuting respectively.
The  ``partitions'' column counts the number of unique circuits required to generate at least one sample of each term in the Hamiltonian.
Gate counts and depths are given in terms of arbitrary 1- or 2-qubit gates restricted to the geometry of 2-qubit gates specified in the connectivity column.
The ``classical cost'' column reports the overhead to determine the partitions for a given Hamiltonian.
In the ``RDM'' column we report whether the technique is able to measure the entire fermionic 2-RDM with the stated scaling, or just a single expectation value (e.g., of the Hamiltonian).
In the ``sym.'' column we report whether any symmetries of the system commute with all measurements made - this allows for simultaneous measurement, enabling strategies for error mitigation by post-selection at zero additional cost.
}
\end{table*}

In this work, we provide schemes for the estimation of fermionic and qubit $k$-RDMs that minimize the number of unique measurement circuits required, significantly decreasing the time required for partial state tomography over prior art.
We demonstrate a scheme to estimate qubit $k$-RDMs in an $N$-qubit system in time ${\cal O}(3^{k}\log^{k-1}\!N)$\footnote{Here and throughout this paper all logarithms are base two.}, achieving an exponential increase over prior art.
We then prove a lower bound of $\Omega(N^k)$ on the number of state preparations required to estimate fermionic $k$-RDMs (such as those of interest in the electronic structure problem) using Clifford circuits (including the addition of ancilla qubits prepared in the $|0\rangle$ state) and measurement in the computational basis.
We describe protocols to achieve this bound for $k\leq 2$.
We detail measurement circuits for these protocols with circuit depths of ${\cal O}(N)$ and gate counts of ${\cal O}(N^2)$ (requiring only linear connectivity), that additionally allow for error mitigation by symmetry verification~\cite{bonet2018error,mcardle2019error}.
Finally, we detail an alternative scheme to measure arbitrary linear combinations of fermionic $k$-RDM elements, based on finding large sets of anti-commuting operators. This requires ${\cal O}(N^4 / \omega)$ measurements, but has a measurement circuit gate count of only ${\cal O}(\omega)$ on a linear array, for a free parameter $\omega < N$.

In Tab.~\ref{tab:comparison}, we provide a history of previous art in optimizing measurement schemes for the electronic structure problem, and include the new results found in this work. We further include the lower bounds for the number of partitions required for anti-commuting and commuting clique cover approaches that were presented in this work.

\section{Background}
Physical systems are characterized by local observables.
However, the notion of locality depends on the exchange statistics of the system in question.
In an $N$-qubit system, data about all $k$-local operators within a state $\rho$ is given by the (qubit) $k$-reduced density matrices, or $k$-RDMs~\cite{rubin2018application}
\begin{equation}
    {}^k\rho_{i_1,\ldots,i_k}=\mathrm{Trace}_{j\neq i_1,\ldots,i_k}[\rho].
\end{equation}
Here, the trace is over all other qubits in the system.
To estimate ${}^k\rho$, we need to estimate expectation values of all tensor products of $k$ single-qubit Pauli operators $P_i\in\{X,Y,Z\}$; we call such tensor products '$k$-qubit' operators.
In an $N$-fermion system, data about all $k$-body operators is contained in the (fermionic) $k$-body reduced density matrices, which are obtained from $\rho$ by integrating out all but the first $k$ particles~\cite{rubin2018application}
\begin{equation}
    {}^kD=\mathrm{Trace}_{k+1,\ldots,N}[\rho].
\end{equation}
Estimating ${}^kD$ requires estimating the expectation values of all products of $k$ fermionic creation operators $c^{\dag}_j$ with $k$ fermionic annihilation operators $c_j$.
For instance, the 2-RDM catalogues all 4-index expectation values of the form $\avg{c^\dagger_p c^\dagger_q c_r c_s}$.
One can equivalently describe fermionic systems in the Majorana basis,
\begin{equation}
    \gamma_{2j}=c_j+c^{\dag}_j,\hspace{0.5cm} \gamma_{2j+1}=i(c^{\dag}_j-c_j),\label{eq:Majoranas}
\end{equation}
in which case the fermionic $k$-RDM may be computed from the expectation values of $2k$ Majorana terms $\gamma_j$ (e.g. the 2-RDM is computed from expectation of Majorana operators of the  form $\avg{\gamma_i \gamma_j \gamma_k \gamma_l}$).
We call such products $2k$-Majorana operators for short.

The expectation values of the above operators may be estimated with standard error $\epsilon$ by $\cal{O}(\epsilon^{-2})$ repeated preparation of $\rho$ and direct measurement of the operator.
This estimation may be performed in parallel for any number of $k$-qubit operators $\hat{P}_i$ or $2k$-Majorana operators $\hat{G}_i$, as long as all operators to be measured in parallel commute.
This suggests that the speed of a `partial state tomography' protocol that estimates expectation values of all $k$-qubit or $2k$-Majorana operators by splitting them into a set of `commuting cliques' (sets where all elements commute) is proportional to the number of cliques required.
In this work we focus on optimizing partial state tomography schemes by minimizing this number.
Necessarily, our approach will be different for qubit systems (where two spatially separated operators always commute) compared to fermionic systems (where this is often not the case).

\section{Near-optimal measurement schemes for local qubit and fermion operators}
Partial state tomography of qubit $k$-RDMs can be efficiently performed by rotating individual qubits into the $X$, $Y$, or $Z$ basis and reading them out.
These rotations define a `Pauli word' $W\in\{X,Y,Z\}^N$, where $W_i$ is the choice of basis for the $i$th qubit.
Repeated sampling of $W$ allows for the estimation of expectation values of any Pauli operator $P$ that is a tensor product of some of the $W_i$ --- we say these operators are contained within the word.
(The set of all such $P$ is the clique corresponding to $W$ with the property that each $P$ is qubit-wise commuting with the rest of operators in the word $W$.)
To estimate the $k$-qubit RDM in this manner, we need to construct a set of words that contain all $k$-local operators.
For $k=2$, it is sufficient to find a set of words $W\in\{A_0,A_1\}^N$ such that each pair of qubits differ in their choice of letter in at least one word.
Then, permuting over $A_0=X,Y,Z$, and separately $A_1=X,Y,Z$, extends the set to contain all $2$-qubit operators.
Such a set can be found via a binary partitioning scheme, for a total of $6\lceil\log N\rceil+3$ cliques (see App.~\ref{app:qubit_schemes} for details).
This scheme may be further extended to arbitrary $k>2$ with a complexity $O(3^k\log^{k-1}(N))$.
The (classical) computational complexity to generate each word is at most $O(\log(N))$, and $O(N)$ to assign each qubit, making the classical computational cost to generate the set of measurements ${\cal O}(e^kN\log^kN)$, which is acceptably small for even tens of thousands of qubits.
We have added code to generate the full measurement protocol to the Openfermion software package~\cite{openfermion}.

Fermionic $k$-RDMs require significantly more measurements to tomograph than their qubit counterparts, as many more operators anti-commute.
In a $N$-fermion system, the total number of $2k$-Majorana operators is ${2N\choose 2k}$, while the size of a commuting clique of $2k$-Majorana operators may be upper-bounded by ${N \choose k}$ in the $N>>k$ limit (see App.~\ref{app:fermionic_limits}).
As fermionic $k$-RDMs contain expectation values of $2k$-Majorana operators, the number of cliques required to estimate all elements in the fermionic $k$-RDM scales as
\begin{equation}
    {2N\choose 2k}/{N \choose k}\sim N^{k}.\label{eq:krdm_bound}
\end{equation}
In terms of the resources requirement to estimate a fermionic $k$-RDM, this directly implies
\begin{thm}
The number of preparations of an arbitrary N-fermion quantum state $\rho$ required to estimate all terms in the fermionic $k$-RDM to within an error $\epsilon$, via Clifford operations (including addition of ancilla qubits prepared in the $|0\rangle$ state), and measurement in the computational basis, is bounded below in the worst case as
$\Omega(\epsilon^{-2}N^k)$.
\label{thm:msmt_bound}
\end{thm}
Proof details may be found in App.~\ref{app:msmt_bound_proof}.
In particular, estimating the fermionic $1$-RDM requires repeated preparation of $\rho$ and measurement over at least $2N-1$ unique commuting cliques, and estimating the fermionic $2$-RDM requires repeat preparation and measurement over a number of cliques at least
\begin{equation}
    \frac{4}{3}N^2-\frac{8}{3}N+1. \label{eq:2rdm_lower_bound}
\end{equation}

Maximally-sized cliques of commuting $2k$-Majorana operators may be achieved via a pairing scheme.
If we pair the $2N$ individual $1$-Majorana operators into $N$ pairs $\{\gamma_i\gamma_j\}$, the corresponding set of operators $i\gamma_i\gamma_j$ forms a commuting set.
Any product of $k$ of these pairs will also commute, so the set of all combinations of $k$ pairs is a commuting clique of exactly ${N\choose k}$ $2k$-Majoranas.
We say that the $2k$-Majorana operators are contained within the pairing.
Curiously, each pairing saturates the bounds found in App.~\ref{app:fermionic_limits} for the number of mutually commuting $2k$-Majorana operators in a $N$-fermion system, and thus this scheme is optimal in the number of $2k$-Majorana operators targeted per measurement circuit.
However, as one $2k$-Majorana operator may be contained in multiple pairings, it remains to find a scheme to contain all $2k$-Majorana operators in the minimum number of pairings.
For the $1$-RDM, it is possible to reach the lower bound of $2N-1$ cliques by a binary partition scheme, which we detail in App.~\ref{app:fermionic_schemes}.
In the $2$-RDM case, we have been able to achieve $\frac{10}{3}N^2$ cliques (also detailed in App.~\ref{app:fermionic_schemes}) by a divide and conquer approach.
It remains an open question whether the factor 5/2 between our scheme and the lower bound (Eq.~\ref{eq:2rdm_lower_bound}) can be improved, either by better bounding or a different scheme.

Simultaneous estimation of the expectation value of each observable may be achieved by repeatedly preparing and measuring states in the $i\gamma_i\gamma_j$ basis for all paired $\gamma_i$, $\gamma_j$ in the clique.
Measuring the system in this basis is non-trivial and depends on the encoding of the fermionic Hamiltonian onto the quantum device.
However, for almost all encodings this requires simply permuting the Majorana labels, which may be achieved by a single-particle basis rotation using Clifford gates (see App.~\ref{app:fermionic_circuits}).
This implies that the circuit depth should be no worse that $O(N)$, and will not require T-gates in a fault-tolerant setting.
Furthermore, in many cases the measurement circuit should be able to be compiled into the state preparation circuit, reducing its cost further.

\begin{figure}
    \centering
    \includegraphics[width=\columnwidth]{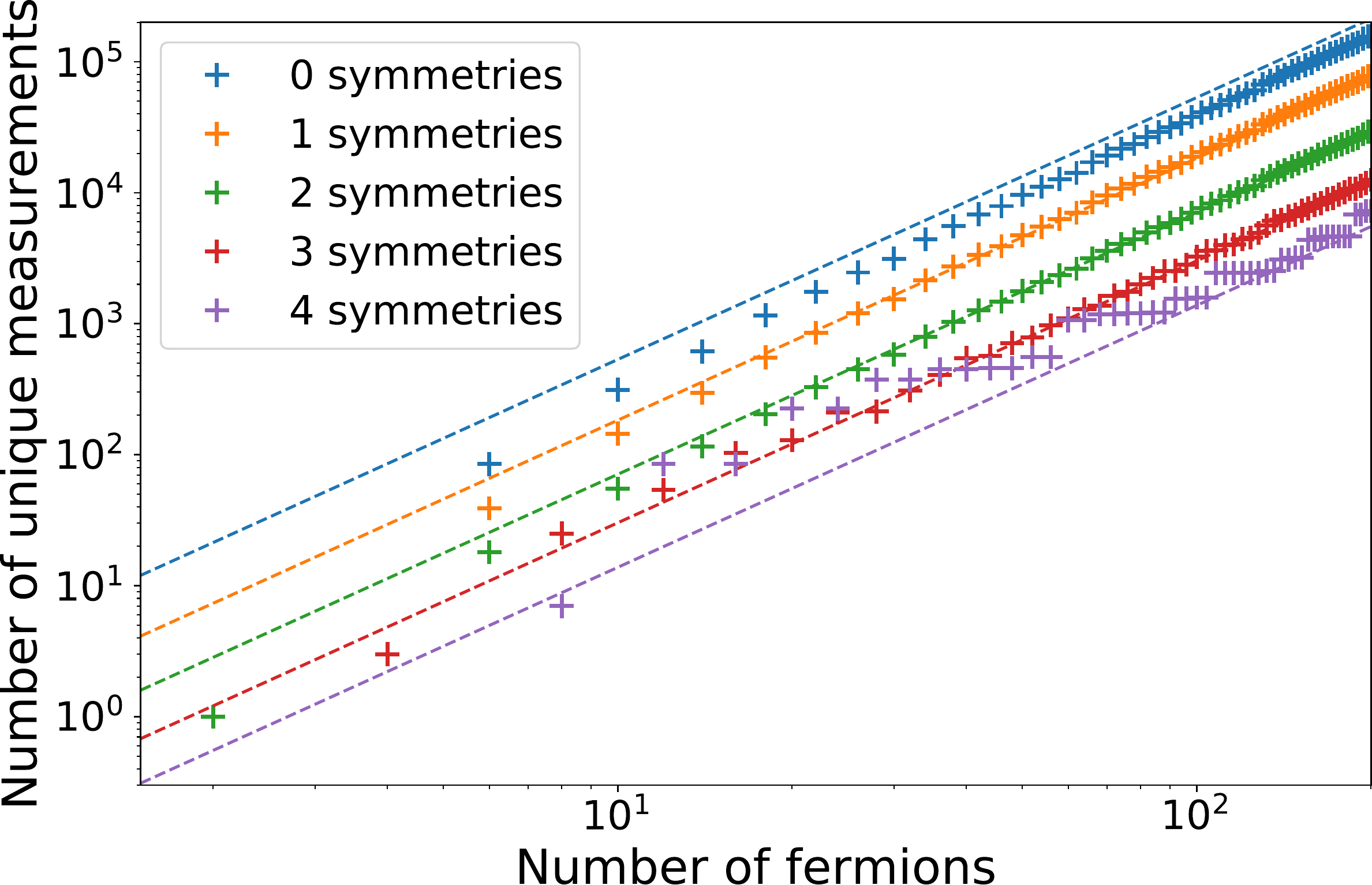}
    \caption{Scaling of our Majorana partitioning scheme in the presence of between $0$ and $4$ symmetry constraints on the system. Dashed lines are from Eq.~\ref{eq:symm_first_order}}
    \label{fig:symm_scaling}
\end{figure}

Symmetry constraints on a system (i.e. unitary or antiunitary operators $S$ that commute with the Hamiltonian $H$) force certain RDM terms to be $0$ for any eigenstates of the system.
For example, when a real Hamiltonian is written in terms of Majorana operators (using Eq.~\ref{eq:Majoranas}), it must contain an even number of odd-index $1$-Majorana operators, and expectation values of terms not satisfying this constraint on eigenstates will be set to $0$.
More generally, if a symmetry is a Pauli word $W_{symmetry}$ such that $W_{symmetry}^2=1$, then it will divide the set of all Majorana terms into those which commute with $W_{symmetry}$ and those which anti-commute; products of odd numbers of anti-commuting terms will have zero expectation value on eigenstates of the system.
Given $n$ such independent symmetries, each of which commute with half of all $1$-Majorana operators (which is typical), we are able to contain all elements of the fermionic $2$-RDM in a number of cliques scaling to first order as 
\begin{equation}
    N^2\left(\frac{10}{3}4^{-N_{\mathrm{sym}}}+2^{1-N_{\mathrm{sym}}}\right).\label{eq:symm_first_order}
\end{equation}
(See App.~\ref{app:symmetries} for details.)
In Fig.~\ref{fig:symm_scaling}, we show the result of an implementation of our scheme for different numbers of symmetries at small $N$, and see quick convergence to this leading-order approximation for up to $4$ symmetries (typical numbers for quantum chemistry problems).
Code to generate this measurement scheme has been added to the OpenFermion package~\cite{openfermion}.

\section{Measuring anti-commuting linear combinations of local fermionic operators}

Products of Majorana and Pauli operators have the special property that any two either strictly commute or strictly anti-commute.
This raises the question of whether there is any use in finding cliques of mutually anti-commuting Pauli operators.
Such cliques may be found in abundance when working with Majoranas --- e.g. for fixed $0\leq j,k,l\leq 2N$, the set $A_{j,k,l}=\{\gamma_i\gamma_j\gamma_k\gamma_l\}$ is a clique of $2N-3$ mutually anti-commuting operators.
Curiously, it turns out that asymptotically larger anti-commuting cliques are not possible - the largest set of mutually anti-commuting Pauli or Majorana operators contains at most $2N+1$ terms (see App.~\ref{app:proof_ac} for a proof). The number of anti-commuting cliques required to contain all $4$-Majorana operators is thus bounded below by $\Omega(N^3)$, matching the numerical observations of~\cite{izmaylov2019unitary}.

Although sampling each term in an anti-commuting clique $A$ of size $L$ requires $\Oo(L)$ state preparations, it is possible to measure a (real) linear combination ${\cal O}=\sum_{i=1}^Lc_i P_i$ of clique elements in a single shot.
Since all elements of $A_{j,k,l}$ share three of the same four indices, here we can associated each $P_i$ in the sum over the elements of $A_{j,k,l}$ with the Majorana $P_i = \gamma_i \gamma_j \gamma_k \gamma_l$.
Given that $\tilde{O}=(\sum_{i=1}^Lc_i^2)^{-1/2}O$ looks like a Pauli operator ($\tilde{O}^{\dag}=\tilde{O}$, $\mathrm{Trace}[\tilde{O}]=0$), and smells like a Pauli operator ($\tilde{O}^2=1$), it can be unitarily transformed to a Pauli operator of our choosing.
In App.~\ref{app:fermionic_circuits}, we show that for systems encoded via the Jordan-Wigner transformation, this unitary transformation may be achieved with a circuit depth of only $N-2+O(1)$ $2$-qubit gates.
It is possible to reduce the depth further by removing Majoranas from the set --- if we restrict ourselves to subsets of $\omega$ elements of $A_{j,k,l}$, the measurement circuit will have $\omega$ gates and be depth $\omega$, but ${\cal O}(N^4/\omega)$ such sets will be needed to estimate arbitrary linear combinations of $4$-Majorana operators.
This makes this scheme very attractive in the near-term, where complicated measurement circuits may be prohibited by low coherence times in NISQ devices.

\section{Conclusion}

Experimental quantum devices are already reaching the stage where the time required for partial state tomography is prohibitive without optimized scheduling of measurements.
This makes work developing new and more-optimal schemes for partial tomography of quantum states exceedingly timely.
In this work, we have shown that a binary partition strategy allows one to sample all $k$-local qubit operators in a $N$-qubit system in poly-$\log(N)$ time, reaching an exponential improvement over previous art.
By contrast, in fermionic systems we have found a lower bound on the number of unique measurement circuits required to directly sample all $k$-local operators of $\Omega(N^{\lceil k/2\rceil})$, an exponential separation.
We have developed schemes to achieve this lower bound for $k=2$ and $k=4$, allowing estimation of the entire fermionic 2-RDM to constant error in ${\cal O}(N^{2})$ time.
Additionally, we have demonstrated that one can leverage the anti-commuting structure of fermionic systems by constructing such sets of size $1\leq \omega\leq N$ to measure all 4-Majorana operators in ${\cal O}(N^{4}/\omega)$ time with a gate count and circuit depth of only $\omega$, allowing one to trade off an decrease in coherence time requirements for an increase in the number of measurements required.
We note that during the final stages of preparing this manuscript a preprint was posted to arXiv which independently develops a similar scheme for measuring $k$-qubit RDMs \cite{cotler2019quantum}.
This scheme seems to be identical to ours for $k=2$ but uses insights about hash functions to generalize the scheme to higher $k$ with scaling of $e^{{\cal O}(k)} \log N$ which improves over our bound of ${\cal O}(3^k \log^{k-1}\!N)$ by polylogarithmic factors in $N$.

\subsection*{Acknowledgments}
We would like to thank  C.W.J. Beenakker, Y. Herasymenko, S. Lloyd, W. Huggins, B. Senjean, V. Cheianov, M. Steudtner, A. Izmaylov, J. Cotler and Z. Jiang for fruitful discussions. This research was funded by the Netherlands Organization for Scientific Research (NWO/OCW) under the NanoFront and StartImpuls programs, and by Shell~Global~Solutions~BV.

\appendix

\section{Schemes for partial state tomography of qubit $k$-RDMs}\label{app:qubit_schemes}

In this section, we develop methods to minimize the measurement cost for partial state tomography of qubit $k$-RDMs by minimizing the number of commuting cliques needed to contain all $k$-qubit operators.
To do so, we associate a `Pauli word' $W\in\{X,Y,Z\}^N$ to each clique: by measuring the $i$th qubit in the $W_i$ basis, we measure every tensor product of the individual Pauli operators $W_i$.
Thus, the clique associated to $W$ contains all $k$-qubit operators that are tensor products of the $W_i$ --- we say these operators are `contained' within the word.
We then wish to find the smallest possible set of words such that every $k$-qubit operator is contained within at least one word.

\begin{figure}
    \centering
    \includegraphics[width=\columnwidth]{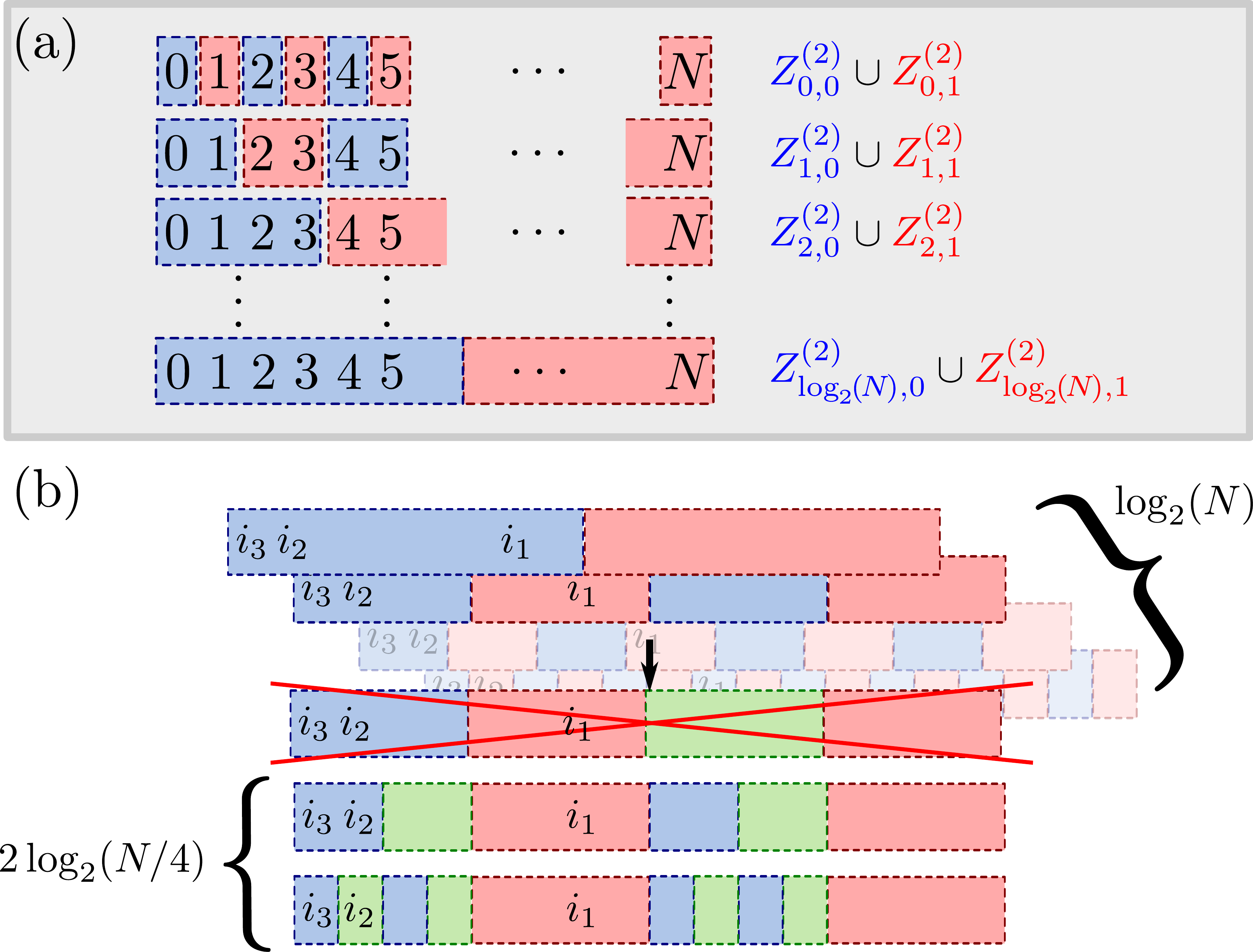}
    \caption{Schematics of the binary partition strategy described in text. (a) Scheme to construct ${\cal O}(\log N)$ cliques that contain all $2$-qubit operators. (b) Extension of the top scheme to a set of ${\cal O}(\log^2 N)$ cliques that contain all $3$-qubit operators.}
    \label{fig:qubit_decomposition}
\end{figure}

We construct such a set through a $k$-ary partitioning scheme, which we first demonstrate for $k=2$.
As motivation, consider that the set of $9$ words (with $A,B=X,Y,Z$)
\begin{equation}
    W^{(A,B)}_i=\begin{cases}A& \mathrm{if}\; i<N/2\\ B&\mathrm{if}\; i\geq N/2 \end{cases},
\end{equation}
contains all $2$-qubit operators that act on qubits $j<N/2$ and $k\geq N/2$.
We may generalize this to obtain all other $2$-qubit operators by finding a set of binary partitions $S_{n,0}\cup S_{n,1}=\{1,\ldots,N\}$ such that for any pair $0\leq i\neq j\leq N$ there exists $n,a$ such that $i\in S_{n,a}$, $j\in S_{n,1-a}$.
Let us define $L=\lceil\log N\rceil$, and write each qubit index $i$ in a binary representation, $i = [i]_{L-1}[i]_{L-2} \ldots [i]_1 [i]_{0}$.
Then, for $n=0,\ldots, L-1$ we define
\begin{equation}
    i\in S_{n,a} \text{ if } [i]_n = a.\label{eq:assigncond}
\end{equation}
All $0\leq i\neq j\leq N$ differ by at least one of their first $L$ binary digits (as shown in Fig.~\ref{fig:qubit_decomposition}(a)), so the set of words $W^{(A,B)}_n$, constructed as
\begin{equation}
    [W^{(A,B)}_{n}],i=\begin{cases}
    A & \mathrm{if} \; i\in S_{n,a}\\
    B & \mathrm{if} \; i\in S_{n,1-a},
    \end{cases}
\end{equation}
defines a set of cliques that contain all $2$-qubit operators.
As $W^{(A,A)}_{n,i}$ is the same word for every $n$ we need only choose this word once and so the number of cliques may be reduced to $6L+3$.

To see how the above may be extended to $k>2$, let us consider $k=3$.
We wish to find $3$-ary partitions $\cup_{a=1}^3S_{n,a}=\{1,\ldots,N\}$ that, given any set $i_1,i_2,i_3$, we can find some index $n$ for which $i_a\in S_{n,a}$ (allowing for permutation of the $i_a$).
Then, by running over all combinations of $X,Y,Z$ on the three parts of each partition, we will obtain a set of words that contain all $3$-qubit operators.
We illustrate a scheme that achieves this Fig.~\ref{fig:qubit_decomposition}(b).
We iterate first over $n=1,\ldots,L$, and find the largest $n$ such that $i_1,i_2$ and $i_3$ are split into two subsets by a binary partition.
(i.e. where $S_{n,a}\cap\{i_1,i_2,i_3\}$ is non-empty for $a=0$ and $a=1$).
This implies that two of the indices lie in one part, and one in the other.
Without loss of generality, let us assume $i_1\in S_{n,1}$ and $i_2,i_3\in S_{n,0}$ (following Fig.~\ref{fig:qubit_decomposition}).
It now suffices to find a set of partitions for $S_{n,0}$ so that we guarantee $i_2$ and $i_3$ are split in one such partition.
We could imagine repeating the binary partition scheme over all $S_{n,0}$; i.e. generating the $\log N$ sets $S_{n,0}\cap S_{n',a}$.
However, we can do better than this.
As $i_1,i_2$ and $i_3$ are not split in any binary partition $S_{n',0},S_{n',1}$ with $n'>n$, $i_2$ and $i_3$ must be in a contiguous block of length $1/2^n$ within $S_{n,0}$.
This means that we need only iterate over $n'=0,\ldots,n-1$.
We must also iterate over the same number of partitions of $S_{n,1}$, and so the total number of partitions we require is
\begin{equation}
    2\sum_{n=0}^{L-1}n=(L-1)(L-2).
\end{equation}

The above generalizes relatively easily to $k>2$.
Given a set $I=\{i_1,\ldots,i_k\}$, we find the binary partition $S_{n,0},S_{n,1}$ with the largest $n$ that splits I into non-empty sets $I_0=I\cap S_{n,0}$ and $I_1=I\cap S_{n,1}$.
Then, we iterate over $|I_0|$-ary partitions of the contiguous blocks of $S_{n,0}$ and the $|I_1|$-ary blocks of $S_{n,1}$.
In total there are $k-1$ possible ways of dividing $I$ (up to permutations of the elements).
This implies that at each $n$ we have to iterate over $k-1$ different sub-partitioning possibilities, making the leading-order contribution to the number of cliques
\begin{equation}
    (k-1)\sum_{n=0}^{L-1}n^{k-2}\sim L^{(k-1)},
\end{equation}
and the total number of cliques ${\cal O}(3^k\log^{k-1}\! N)$.

\section{Upper bounds on the size of commuting cliques of Majorana operators.}\label{app:fermionic_limits}

In this appendix, we detail the bounds on the size of commuting cliques of Majorana operators.
Let us call the largest number of mutually-commuting $k$-Majoranas that are a product of $l$ unique terms (i.e. $l$ unique $1$-Majoranas) $M^{k}_l$.
(For an $N$-fermion system, we will eventually be interested in the case where $l=2N$.)
We wish to bound this number $M^k_l$ by induction.
All $1$-Majorana operators anti-commute, so $M^k_l=1$.
Then, let us consider the situation where $k$ is even and when $k$ is odd separately.
Suppose we have a clique of $M^k_l$ $k$-Majorana operators with $k$ even.
As there are only $l$ unique terms, and these $k$-Majoranas contain $kM^k_l$ individual terms each, there must be a clique of $\lceil kM^k_l/l\rceil$ of these operators that share a single term $\gamma_0$.
We may write each such operator in the form $\pm\gamma_0\Gamma_i$, where $\Gamma_i$.
As $[\gamma_0\Gamma_i,\gamma_0\Gamma_j]=0$ if and only if $[\Gamma_i,\Gamma_j]=0$, this gives a clique of $kM^k_l/l$ commuting $(k-1)$-Majorana operators on $l-1$ unique terms, so we must have
\begin{equation}
\left\lceil\frac{kM^{k}_l}{l}\right\rceil\leq M^{k-1}_{l-1},\hspace{1cm} \mathrm{k\;even}.
\end{equation}
Now, consider the case where $k$ is odd, let us again assume we have a clique of $M^{(k)}$ commuting $k$-Majoranas.
Two products of Majorana operators anticommute unless they share at least one term in common, so let us choose one $k$-Majorana $\Gamma$ in our set; each $k$-Majorana must have at least one of the $k$ terms in $\Gamma$, so at least one such term is shared between $\lceil M^{(k)}/k\rceil $ Majoranas in our set.
Removing this term gives a clique of $\lceil M^{(k)}/k\rceil $ $(k-1)$-Majorana operators on $l-1$ unique terms, and so we have
\begin{equation}
    \left\lceil \frac{M^{k}}{k}\right\rceil \leq M^{k-1}_{l-1},\hspace{1cm} \mathrm{k\;odd}.
\end{equation}
These equations may be solved inductively to lowest-order in $k$ to obtain 
\begin{equation}
M^{k}_l\sim l^{\lfloor k/2\rfloor}.
\end{equation}

This bound can be strengthened in the $l>>k$ limit, as here the largest commuting cliques of odd-$k$-Majoranas must share a single term $\gamma_0$.
This can be seen as when $k$ is odd, large sets of commuting ($k-1$)-Majoranas contain many operators that do not share any terms --- a set of $k-1$ commuting operators that share a single term can be no larger than approximately $l^{(k-3)/2}$.
Formally, let us consider a set $C$ of commuting $k$-Majoranas, choose $\Gamma\in C$, and write $\Gamma=\gamma_1\ldots\gamma_k$.
Then, we may write $C=\cup_i C_i$, where $C_i$ is the subset of operators in $C$ that contain $\gamma_i$ as a term.
If there exists $\Gamma'\in C/C_i$, (i.e. $\Gamma'$ commutes with all operators in $C_i$ but does not itself contain $\gamma_i$), we may divide $C_i$ into $k$ subsets of Majoranas that share the individual terms in $\Gamma'$, and so $|C_i|\leq kl^{(k-3)/2}$.
If is true for all such $C_i$, we have then $|C|\leq \sum_i|C_i|\leq k^2l^{(k-3)/2}$.
As this scales suboptimally in the large-$l$ limit~\footnote{For example, we can achieve better scaling in $l$ via our pairing scheme.}, we must have that $C/C_i$ is empty for some $C_i$.
Then, $C_i=C$, and we can bound
\begin{equation}
M^{k}_l \leq M^{k-1}_{l-1},\hspace{1cm}\mathrm{k\; odd}.\label{eq:tight_odd_bound}
\end{equation}
This leads to the tighter bound (assuming $l$ even)
\begin{equation}
M^{k}_l \leq \frac{l!!}{(l-k)!!\;k!!},
\end{equation}
where the double factorial implies we multiple only the even integers $\leq k$.
Then, when $l=2N$, for even $k=2n$ we see
\begin{align}
M^{2n}_{2N} &\leq \frac{2N!!}{(2N-2n)!!\;2n!!}\nonumber\\&=\frac{2^NN!}{2^{N-n}(N-n)!2^nn!}={N \choose n}.
\end{align}
This is precisely the size of the cliques obtained by pairing, proving this scheme optimal in the large-$N$ limit.

In practice, we observe that Eq.~\ref{eq:tight_odd_bound} is true for $k=3$ whenever $l\geq l_{\mathrm{crit},3}15$ (i.e. for $>8$-fermion systems).
This is because the largest set of commuting $3$-Majoranas that do not share a single common element can be found to be (up to relabeling) $\{\gamma_0\gamma_1\gamma_2,\gamma_0\gamma_3\gamma_4,\gamma_0\gamma_5\gamma_6,\gamma_1\gamma_3\gamma_5,\gamma_1\gamma_4\gamma_6,\gamma_2\gamma_3\gamma_6,\gamma_2\gamma_4\gamma_5\}$, which contains $7$ terms.
The above argument implies that $l_{\mathrm{crit},k}$ scales at worst as $k^2$, however the bounds obtained here are rather loose, and we expect it to do far better.

\section{Details of measurement schemes for fermionic systems}~\label{app:fermionic_schemes}

\begin{figure}
    \centering
    \includegraphics[width=\columnwidth]{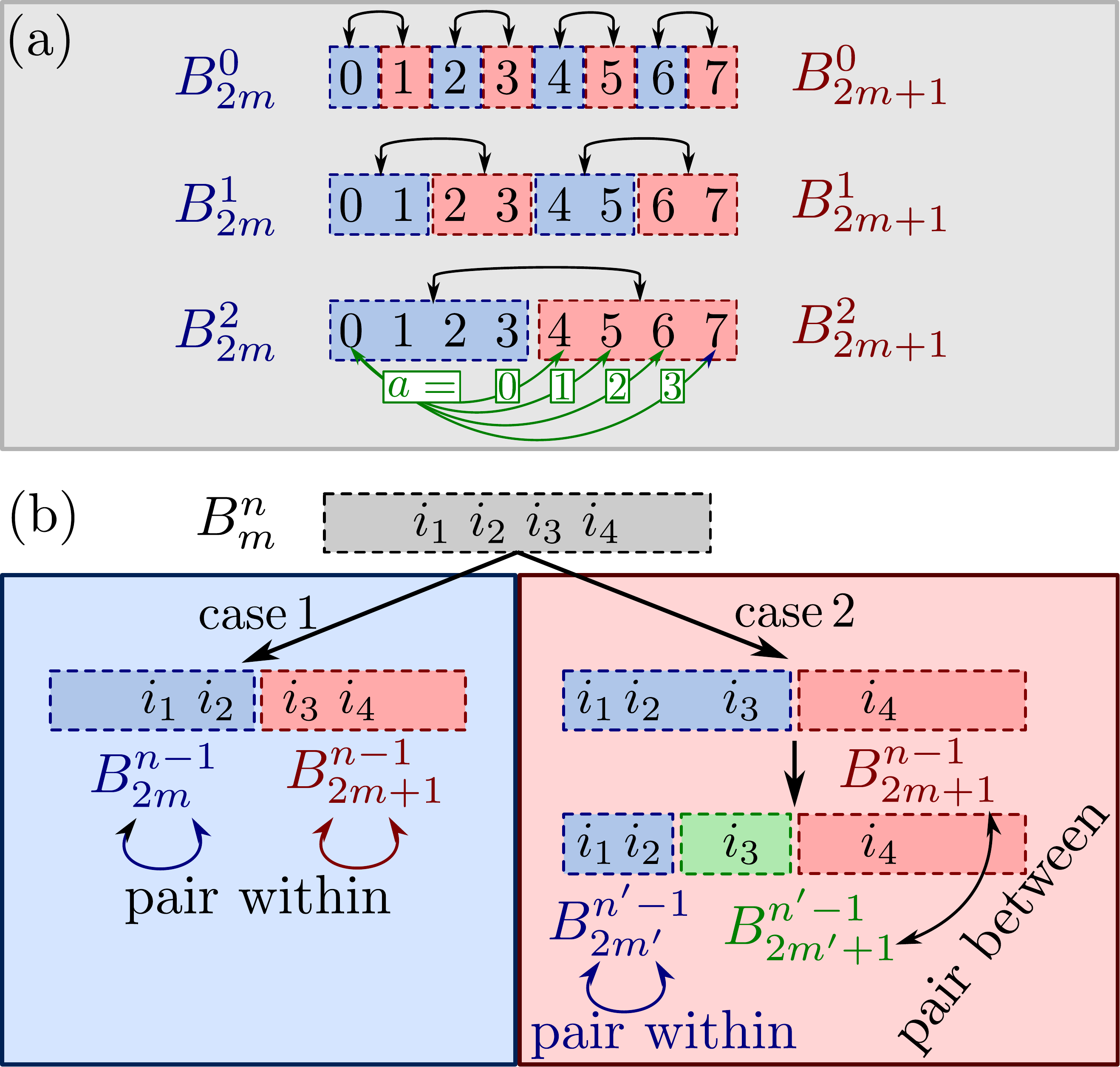}
    \caption{Schematic of the fermionic partition strategy for generating cliques that contain all local fermionic operators. (a) a scheme to pair all indices in $\{1,\ldots,N\}$ in ${\cal O}(N)$ timesteps. (b) The two cases to consider in our strategy to contain all $4$-Majorana operators in only ${\cal O} (N^2)$ cliques.}
    \label{fig:Majorana_partitions}
\end{figure}

We now construct asymptotically minimal sets of cliques that contain all $2$-Majorana and $4$-Majorana operators.
$2$-Majorana operators that share any term do not commute, so our commuting cliques of $2$-Majorana operators must contain only non-overlapping pairs of Majorana terms.
Equivalently, we need to find a set of pairings of $\{0,\ldots,2N\}$ such that each pair $(i,j)$ appears in at least one pairing.
This may be achieved optimally for $N$ a power of $2$ via the partitioning scheme outlined in Fig.~\ref{fig:Majorana_partitions}(a).
We first split $\{1,\ldots,2N\}$ into a set of $N2^{-n}$ contiguous blocks for $n=0,\ldots,\log(2N)$
\begin{equation}
    B^n_m=\{ m\times 2^n \leq i < (m+1)\times 2^n \}.
\end{equation}
Then, our cliques may be constructed by pairing the $i$th element of $B^n_{2m}$ with the $(i+a)$th element of $B^n_{2m+1}$ (modulo $2^n$), as $n$ runs over $0,\ldots,\log (N)$ and $a$ runs over $0,\ldots,2^n-1$.
Formally, this gives the set of cliques
\begin{align}
    C_{a,n}:=\{&\gamma_{\alpha}\gamma_{\beta}, \alpha=(m2^{n+1}+i),\nonumber\\
    &\beta=((2m+1)2^n + [(i+a)\;\mathrm{mod}\;2^n]),\nonumber\\
    &m=(0,\ldots,N2^{-n}-1), i=(0,\ldots,2^{n}-1)\},
\end{align}
with a total number
\begin{equation}
    \sum_{n=0}^{\log N}2^n=2N-1,
\end{equation}
matching exactly the lower bound calculated in the main text.
The above technique needs slight modification when $N$ is not a power of $2$ to make sure that when $|B^n_{2m}|\neq |B^n_{2m+1}|$, unpaired elements are properly accounted for, but the above optimal scaling may be retained.
Code to generate an appropriate set of pairings has been added to the openfermion package~\cite{openfermion}.

As all operators in one of the above cliques $C_{a,n}$ commute, their products commute, and the set 
\begin{equation}
\{\gamma_i\gamma_j\gamma_k\gamma_l;\;\gamma_i\gamma_j,\gamma_k\gamma_l\in C_{a,n}\},
\end{equation}
is clearly a clique of commuting $4$-Majorana operators.
However, each $2$-Majorana operator is guaranteed to be in only one of the cliques $C_{a,n}$, so this will not yet contain all $4$-Majorana operators.
To fix this, we aim to construct a larger set $\{C_{\alpha}\}$ of cliques of commuting $2$-Majorana operators, such that for every set $\gamma_{i_1},\gamma_{i_2},\gamma_{i_3},\gamma_{i_4}$ there exists one $C_{\alpha}$ containing both $\gamma_{i_a}\gamma_{i_b}$ and $\gamma_{i_c}\gamma_{i_d}$ (for some permutation of $a,b,c,d=1,2,3,4$).
This may be achieved by the strategy illustrated in Fig.~\ref{fig:Majorana_partitions}(b).
For each $I={i_1,i_2,i_3,i_4}$, choose the smallest $n$ such that $I\subset B^n_m$ for some $m$.
This implies that the $\{B^n_m\}$ split $I$ into two parts - $I_a=I\cap B^{n-1}_{2m+a}$, for $a=0,1$, and $|I_0|=1,2$ or $3$.
Suppose first $|I_0|=2$, (case 1 in Fig.~\ref{fig:Majorana_partitions}(b)).
In this case, by iterating over all pairs of elements in $B^{n-1}_{2m}$ and subsequently all pairs of elements in $B^{n-1}_{2m+1}$, we will at some point simultaneously pair the elements of $I_0$ and the elements of $I_1$, as required.
This may be performed in parallel for each $m$, making the total number of cliques generated at each $n$ $|B^{n-1}_{2m}|^2=4^{n-1}$.
Now, suppose $|I_0|=3$ (case 2 in Fig.~\ref{fig:Majorana_partitions}) --- or $|I_0|=1$ as the two situations are equivalent.
Let $n'<n$ be the smallest number such that $I_0\subset B^{n'}_{m'}$ for some $m'$, and we may split $I_0$ into two sets $I_{0,a}=I_0\cap B^{n'-1}_{2m'+a}$ for $a=0,1$.
Of the three elements in $I_0$, two of them must either lie in $I_{0,0}$ or $I_{0,1}$ - suppose without loss of generality that $|I_{0,0}|=2$.
Then, by iterating over all pairs within $B^{n'-1}_{2m'}$, and all pairs between elements of $B^{n'-1}_{2m'+1}$ and $B^{n-1}_{2m+1}$, we will at some point pair both elements in $I_{0,0}$ and both elements in $I_{0,1}\cup I_1$.

This pairing needs to occur for all $n>n'$, which implies we need to iterate over all combinations of pairs between elements of $B^{n'-1}_{2m'+1}$ and $\{1,\ldots,2N\}/B^{n'-1}_{2m'}$ (while iterating over pairs within $B^{n'-1}_{2m'}$).
This may be performed in parallel for each $m'$ at each $n'$.
First, iterate over all possible pairings of $B^{n'}_{m_0}$ and $B^{n'}_{m_1}$ (which requires ${\cal O}(N2^{-n'})$ iterations).
Then, iterate over all pairs between $B^{n'-1}_{2m_0+a_0}$ and $B^{n'-1}_{2m_1+a_1}$ for all combinations of $a_0,a_1=0,1$ (requiring $4\times 2^{n'-1}$ iterations).
Simultaneously, iterate over all pairs within $B^{n'-1}_{2m_0+1-a_0}$ and $B^{n'-1}_{2m_1+1-a_1}$ (requiring again $2^{n'-1}$ iterations).
This generates $4\times 4^{n'-1}$ cliques at each $n'$.
The total number of cliques we then require to contain all $4$-Majorana operators using this scheme is then
\begin{equation}
    \sum_{n'=1}^{\lceil\log N\rceil}N2^{n'}+\sum_{n=1}^{\lceil\log N\rceil+1}4^{n-1}\sim\frac{10}{3}N^2.
\end{equation}

\section{Reducing operator estimation over symmetries}\label{app:symmetries}
Given a set $\{S_i\}\subset\PP^N$ of $N_{\mathrm{sym}}$ mutually-commuting Pauli operators that are symmetries ($[S_i,H]=0$), we can simultaneously diagonalize both the Hamiltonian and the symmetries, implying that we can find a ground state $\rho$ such that $\mathrm{Trace}[\rho P]=0$ for each $P$ that does not commute with $S_i$.
In the case of a degenerate ground state eigenspace, not all states will necessarily have this property (as symmetries may be spontaneously broken).
However, any such $P$ will not appear in the Pauli decomposition of the Hamiltonian, and so estimation of this RDM term is not necessary to calculate the energy of the state.
The commutation of a $k$-Majorana operator $\Gamma$ with a Pauli operator symmetry $S_i$ may be seen immediately by counting how many of the $k$ individual terms anti-commute with $S_i$ --- if this number is even, then $[\Gamma,S_i]=0$.
This implies that we can separate individual $1$-Majorana operators into bins $B_{\vec{s}}$ with $\vec{s}\in\{0,1\}^{N_{\mathrm{sym}}}$ a commutation label:
\begin{equation}
    \gamma_j\in B_{\vec{s}}\rightarrow \begin{cases} \gamma_jS_i=S_i\gamma_j & \mathrm{iff}\; s_i = 0 \\ \gamma_jS_i=-S_i\gamma_j & \mathrm{iff} \; s_i = 1.
    \end{cases}
\end{equation}
Let $\vec{s}(\gamma_j)$ denote the label of the bin $\gamma_j$ may be found in, and we may generalize to all $k$-Majorana operators $\Gamma=\prod_{l=1}^k\gamma_{j_l}$:
\begin{equation}
    s_i\left(\prod_{l=1}^k\gamma_{j_l}\right)=\sum_l s_i(\gamma_{j_l}) \mod 2.
\end{equation}
To estimate the symmetry-conserved sector of the $2$-RDM, we are then interested in constructing a set of cliques of $4$-Majorana operators in $B_{\vec{0}}$.
These take the form $\gamma_{j_1}\gamma_{j_2}\gamma_{j_3}\gamma_{j_4}$ where $\vec{s}(\gamma_{j_1})=\vec{s}$, $\vec{s}(\gamma_{j_2})=\vec{s}+\vec{\delta}$, $\vec{s}(\gamma_{j_3})=\vec{s}+\vec{\alpha}$, and $\vec{s}(\gamma_{j_4})=\vec{s}+\vec{\alpha}+\vec{\delta}$.
(Recall here that in binary vector arithmetic, $\vec{a}+\vec{a}\mod 2=\vec{0}$.)
We construct cliques for the above in two steps.
First, we iterate over all quadruples within each bin $B_{\vec{s}}$ (using the methods in App.~\ref{app:fermionic_schemes}).
This covers all of the above operators where $\vec{\delta}=\vec{\alpha}=0$, and may be done simultaneously with cost $\frac{10}{3}B^2$, where $B$ is the size of the largest bin.
Then, we iterate between bins $B_{\vec{s}}$ and $B_{\vec{s}+\vec{\beta}}$ for all $\beta\in \{0,1\}^{N_{\mathrm{sym}}}$ with $\beta_0=0$.
Such iteration achieves all pairs above --- either $\alpha_0=0$ (and we pair bins $B_{\vec{s}}$ with $B_{\vec{s}+\vec{\alpha}}$ when we pair $B_{\vec{s}+\vec{\delta}}$ with $B_{\vec{s}+\vec{\delta}+\vec{\alpha}}$), or $\delta_0=0$ (and we pair bins $B_{\vec{s}}$ with $B_{\vec{s}+\vec{\delta}}$ when we pair $B_{\vec{s}+\vec{\alpha}}$ with $B_{\vec{s}+\vec{\delta}+\vec{\alpha}}$), or $(\vec{\delta}+\vec{\alpha})_0=0$ (and we pair $B_{\vec{s}}$ with $B_{\vec{s}+\vec{\delta}+\vec{\alpha}}$ when we pair $B_{\vec{s}+\vec{\alpha}}$ with $B_{\vec{s}+\vec{\delta}}$).
We must perform this pairing in parallel - i.e. construct a set of $2^{N_{\mathrm{sym}}-1}$-tuples by drawing one element from each $B_{\vec{s}}\times B_{\vec{s}+\delta}$ such that every two elements appear in at least one tuple.
In App.~\ref{app:parallel_iterate} we describe how this may be achieved
The total cost of the above is then $2^{N_{\mathrm{sym}}-1}(B^2+2B\ln(B)+\ln(B)^2)$.
It is common for most symmetries to divide the set of Majoranas in two, in which case $B=2N\times 2^{-N_{\mathrm{sym}}}$, and our clique cover size is
\begin{equation}
    N^2\left(\frac{10}{3}4^{-N_{\mathrm{sym}}}+2^{1-N_{\mathrm{sym}}}\right) + O(N\ln(N)).
\end{equation}
We summarize our method in algorithm~\ref{alg:sym_iter} (where we use $h(\vec{s})$ as the Hamming weight of a binary vector $\vec{s}$).

\begin{algorithm}[H]
\label{alg:sym_iter}
\caption{Iterate over symmetry-conserved $2$-RDM elements. Here, iterQuad and pairBetween are described in App.~\ref{app:fermionic_schemes}, and parallelIterate in Alg.~\ref{alg:parallel_iter}}
\begin{algorithmic}
\State Construct bins $B_{\vec{s}}$.
\State quadIter = $\{\}$
\For {$\vec{s}$ in $\{0,1\}^{N_{\mathrm{sym}}}$}
    \State quadIter[$\vec{s}$] = iterQuad($B_{\vec{s}}$)
\EndFor
\While {any iterator in quadIter is not stopped}
    yield (next(iterator) for iterator in quadIter if iterator is not stopped)
\EndWhile
\State seriesIterate(quadIter)
\For {$\vec{\beta}$ in $\{0,1\}^{N_\mathrm{sym}-1}$, $\beta\neq \vec{0}$}
    \State Left-append $0$ to beta (i.e. $\beta = (0,) + \beta$)
    \State quadIter = $\{\}$
    \For {$\vec{s}$ in $\{0,1\}^{N_{\mathrm{sym}}}$, $h(\vec{s}+\vec{\beta})\geq h(\vec{s})$}
        \State quadIter[$\vec{s}$] = pairBetween($B_{\vec{s}}$., $B_{\vec{s}+{\vec{\beta}}})$.
    \EndFor
    \State parallelIterate(quadIter)
\EndFor
\end{algorithmic}
\end{algorithm}

\section{Parallel iteration over pairings}\label{app:parallel_iterate}
If we wish to iterate over all pairs of two lists of $L$ elements each, clearly we must perform at least $L^2$ total iterations, and the optimal strategy is trivial (two loops).
However, if we wish to iterate over all pairs between $K=3$ or more lists of $L$ elements (i.e. generate a set of $K$-tuples such that each pair appears as a subset of one tuple), such an optimal strategy is not so obvious.
When $K$ is less than the smallest factor of $L$, a simple algorithm works as described in Algorithm~\ref{alg:parallel_iter}.
We can see that this algorithm works, for suppose $jk_1+l=a\mod L$ and $jk_2+l=b\mod L$ for two separate values of $j, l$ - i.e. $j_1k_1+l_1=j_2k_1+l_2\mod L$ and $j_1k_2+l_1=j_2k_2+l_2\mod L$. Then, we have $j_1(k_1-k_2)=j_2(k_1-k_2)\mod L$, and as $k_1, k_2$ are smaller than the lowest factor of $L$, $gcd(k_1-k_2,L)=1$, implying $j_1=j_2$.
This scheme achieves the optimal $L^2$ total iterations, although the reliance on $K$ being smaller than the lowest factor of $L$ is somewhat unsavoury.
We hypothesize that the asymptotic $L^2$ is indeed achievable for all $K\leq L$, but have not been unsuccessful in our search for a construction.
Instead, for composite $L$, we suggest padding each list to have length $L'$, being the first number above $L$ that achieves this requirement.
The prime number theorem implies that $L'-L\sim\log(L)$ if $K\leq L$ (as then we require at worst to find the next prime number).
This gives the scheme runtime $L^2+2L\log(L)+\log(L)^2$, which is a relatively small subleading correction.

\begin{algorithm}[H]
\caption{parallelIterate: Iterate over $K$ lists $dataArray[0],...,dataArray[K-1]$ of $L$ elements, generating all pairs between elements in separate lists. Assumes $K$ less than the smallest factor of $L$.}\label{alg:parallel_iter}
\begin{algorithmic}
\For {$j=0$ to $L-1$}
    \For {$k=0$ to $L-1$}
        \State thisTuple = [dataArray[$k$][$jk+l\mod L$] for $k=0$ to $K-1$]
        \State yield thisTuple
    \EndFor
\EndFor
\end{algorithmic}
\end{algorithm}

\section{Measurement circuitry for fermionic RDMs}\label{app:fermionic_circuits}

Direct measurement of products of Majorana operators is a more complicated matter than measurement of Pauli words (which require only single-qubit rotations).
However, when the fermionic system is encoded on a quantum device via the Jordan-Wigner transformation~\cite{jordan1928p}, a relatively easy measurement scheme exists.
Within this encoding, we have
\begin{equation}
    i\gamma_{2n}\gamma_{2n+1}=Z_n,
\end{equation}
so if we can permute all Majorana operators such that each pair $(\gamma_i,\gamma_j)$ of Majoranas within a given clique is mapped to the form $(\gamma_{2n},\gamma_{2n+1})$, they may be easily read off.
To achieve such a permutation, we note that the Majorana swap gate $U_{i,j}=e^{\frac{\pi}{4}\gamma_i\gamma_j}$ satisfies
\begin{equation}
    U_{i,j}^{\dag}\gamma_k U_{i,j}=\begin{cases}\gamma_k & \mathrm{if}\; i,j\neq k\\
    \gamma_j & \mathrm{if}\; k=i\\
    -\gamma_i & \mathrm{if}\; k=j.
    \end{cases}
\end{equation}
And so repeated iteration of these unitary rotations may be used to 'sort' the Majorana operators into the desired pattern.
This may be performed in an odd-even search format~\cite{haberman1972parallel} - at each step $t=1,\ldots,N$ we decide for each $n=1,\ldots N$ whether to swap Majoranas $2n$ and $2n+1$, and then whether to swap Majoranas $2n$ and $2n-1$.
Within the Jordan-Wigner transformation these gates are local:
\begin{equation}
    U_{2n,2n+1}=e^{-i\frac{\pi}{4}Z_n},\hspace{0.5cm}U_{2n-1,2n}=e^{-i\frac{\pi}{4}Y_{n-1}Y_n},
\end{equation}
and so each timestep is depth $3$, for a total maximum circuit depth of $3N$ and total maximum gate depth $3N^2$.
(To see that only $N$ timesteps are necessary, note that each Majorana can travel up to $2$ positions per timestep.)
Following the Majorana swap circuit, all pairs of Majoranas that we desire to measure will be rotated to neighbouring positions and may then be locally read out.
As each Majorana swap gate commutes with the global parity $\prod_{i=1}^{2N}\gamma_i$, this will be measurable alongside the clique as the total qubit parity $\prod_{i=1}^{N}Z_i$, allowing for error mitigation by symmetry verification~\cite{bonet2018error,mcardle2019error}.
As the above circuit corresponds just to a basis change, for many VQEs it may be pre-compiled into the preparation itself, negating the additional circuit depth entirely.

As an alternative to the above ideas, it is possible to extend the paritioning scheme for measuring all $k$-qubit operators to a scheme to sample all fermionic 2-RDM elements via the Bravyi-Kitaev transformation \cite{bravyi2002fermionic,Seeley2012}. 
This transformation maps local fermion operators to $k= {\cal O}(\log N)$ qubit operators, and so using our approach the resulting scheme would require ${\cal O}(3^k \log^{k-1}\!N) = (3\log N)^{{\cal O}(\log N)} $ unique measurement. Although this is superpolynomial, it is a slowly growing function for small $N$ and also has the advantage that the measurement circuits themselves are just single qubit rotations. Furthermore, as the set of fermion operators is very sparse in the sense that it has only ${\cal O}(N^4)$ terms rather than $N^{{\cal O}(\log N)}$ terms, the scheme may be able to be further sparsified.

The measurement scheme to transform a sum of anti-commuting Majorana operators to a single Majorana operator follows a similar scheme to the Majorana swap network, but with the swap gates replaced by partial swap rotations.
Let $A$ be a set of anti-commuting Majorana (or Pauli) operators, and then for $P_i,P_j\in A$ the (anti-Hermitian) product $P_iP_j$ commutes with every element in $A$ but $P_i$ and $P_j$ itself.
This implies that the unitary rotation $e^{\theta P_iP_j}$ may be used to rotate between $P_i$ and $P_j$ without affecting the rest of $A$:
\begin{equation}
    e^{-\theta P_iP_j}P_k e^{\theta P_iP_j}=\begin{cases} P_k & \mathrm{if}\; k\neq i,j\\
    \cos(\theta)P_i + \sin(\theta)P_j & \mathrm{if}\; k=i\\
    \cos(\theta)P_j - \sin(\theta)P_j & \mathrm{if}\; k=j
    \end{cases}.
\end{equation}
This rotation may be applied to remove the support of $O$ on individual $P_i$.
For example, if $\theta_{1}=\tan^{-1}(\frac{c_1}{c_2})$
\begin{equation}
    e^{-\theta_1 P_1P_2}O e^{\theta_1 P_1P_2}=\sqrt{c_1^2+c_2^2}P_2+\sum_{i=3}^Lc_iP_i.
\end{equation}
We extend this to remove support of $O$ on each $P_i$ in turn by choosing $\theta_i=\sqrt{\sum_{j<i}c_i^2}/c_{i+1}$, and then
\begin{equation}
    \left(\prod_{i=1}^{L-1}e^{-\theta_i P_iP_{i+1}}\right)O\left(\prod_{i=1}^{L-1}e^{\theta_i P_iP_{i+1}}\right) = \sqrt{\sum_i c_i^2} P_L.
\end{equation}
Following this measurement circuit, $O$ may be measured by reading all qubits in the basis of the final Pauli $P_L$.
Intriguingly, for $P_i, P_{i+1} \in A_{j,k,l}$, we have that $P_iP_{i+1}=\gamma_i\gamma_{i+1}$, which maps to a 2-qubit operator under the Jordan-Wigner transformation (as noted previously).
This implies a measurement circuit for these sets may be achieved with only linear gate count and depth, linear connectivity, and no additional ancillas. We can slightly reduce the depth by simultaneously removing the $P_i$ from the ``top'' and ''bottom''; i.e., we remove $P_{2N-3}$ by rotating with $P_{2N-4}$ at the same time as removing $P_1$ by rotating with $P_2$, until after exactly $N-2$ layers, we have only the term $P_N$ remaining.
All generators in this unitary transformation commute with the parity $\prod_{i=1}^{2N}\gamma_i$, implying that it remains invariant under the transformation and may be read out alongside $P_{N}$.
(This may require an additional ${\cal O}(1)$ gates if $P_N$ is not mapped to products of $Z_i$ via the Jordan-Wigner transformation.)

\section{Proof that the maximum size of an anti-commuting clique of Pauli or Majorana operators is $2N+1$}\label{app:proof_ac}
We prove this result in general for the Pauli group $\PP^N$, and note that as the Jordan-Wigner transformation maps Majorana operators to single elements of $\PP^N$, the same is true of this.
We first note that elements within an anti-commuting clique $S\subset\PP^N$ may not generate each other - let $\prod_{i=1}^n P_i=P_j\in S$, and if $n$ is odd $[P_i,P_j]=0$ for any $P_i$ in the product, while if $n$ is even $[P_k,P_j]=0$ for any $P_k$ not in the product.
(The one exception to this rule is if one cannot find any such $P_k$, i.e. when $P_j=\prod_{i\neq j,P_i\in S}P_i$).
Then, note that each element $P\in\PP^N$ commutes with precisely half of $\PP^N$, and anticommutes with the other half.
This can be seen because a Clifford operation $C$ exists such that $C^{\dag}PC=Z_1$, which commutes with all operators of the form $I_1P'$ and $Z_1P'$ and anti-commutes with all operators of the form $X_1P'$ and $Y_1P'$, and these will be mapped to other Pauli operators when the transformation is un-done.

We may extend this result: a set $S=\{P_1,\ldots,P_n\}$ of $n$ non-generating anti-commuting elements in $\PP^N$ splits $\PP^N$ into $2^n$ subsets $\Pp_{\vec{b}}$ (with $\vec{b}\in\ZZ_2^n$), where $Q\in\Pp_{\vec{b}}$ commutes with $P_i$ if $b_i=0$ (and anticommutes if $b_i=1$).
To see that all $\Pp_{\vec{b}}$ must be the same, note that given an operator $Q\in\Pp_{\vec{b}}$, $P_iP_jQ\in\Pp_{\vec{b}\oplus \vec{\delta}_i\oplus\vec{\delta}_j}$ (as $P_iP_j$ anti-commutes with $P_i$ and $P_j$ but commutes with all other elements in $S$), so $|\Pp_{\vec{b}}|$ and $\Pp_{\vec{b}\oplus \vec{\delta}_i\oplus\vec{\delta}_j}$ are the same size.
Similarly, if $Q\in\Pp_{\vec{b}}$, $P_iQ\in\Pp_{\vec{b}\oplus\vec{1}\oplus\vec{\delta}_i}$.
If $n$ is even, this is sufficient to connect each element in $\Pp_{\vec{b}}$ to an element in $\Pp_{\vec{b}'}$, forcing all to be the same size.
However, if $n$ is odd the above will not connect $\Pp_{\vec{b}}$ and $\Pp_{\vec{b}'}$ unless $|\vec{b}|=|\vec{b}'|\;\mathrm{mod}\;2$. We note that $\bigcup_{\vec{b},|\vec{b}|\;\mathrm{mod}\;2=0}\Pp_{\vec{b}}$ is the set of elements that commute with $\prod_{P_i\in S}P_i$, and thus must be precisely half of $\PP^N$.
This proves that the set of operators in $\PP^N$ that anticommute with all elements in $S$ is of size $4^N/2^{|S|}$.
This must be an integer, so $n\leq 2N$.
Then, when $n=2N$ there is precisely one element that anticommutes with all operators in $S$ - $\prod_{P_i\in S}P_i$, and we may add this to $S$ to get the largest possible set of operators.
Such a set is unitarily equivalent to the set of $2N$ Majorana operators $\gamma_i$ and the global parity $\prod_{i=1}^{2N}\gamma_i$.

\section{Proof of Thm.~\ref{thm:msmt_bound}}\label{app:msmt_bound_proof}
To bound the number of preparations of a state $\rho$ required to estimate a fermionic $k$-RDM, we first establish a correspondence between the allowed measurement protocols and measurement of a set of commuting Pauli operators on the original state $\rho$.
As $2k$-Majorana operators are Pauli operators, this implies that an estimate of the expectation value $\langle\Gamma_i\rangle$ of each $2k$-Majorana operator $\Gamma_i$ converges with variance
\begin{equation}
    \mathrm{Var}(\langle\Gamma_i\rangle)\leq\frac{(1-\langle\Gamma_i\rangle)(1+\langle\Gamma_i\rangle)}{4M_i},\label{eq:var_bound}
\end{equation}
where $M_i$ is the number of preparations and measurements of $\rho$ in a basis containing $\Gamma_i$.
We then show the existence of a worst-case state for which this upper bound is tight, which implies that to estimate $\langle\Gamma_i\rangle$ with error $\epsilon$ we require $M_i\sim\epsilon^{-2}$ preparations and measurements of $\rho$ in a basis containing $\Gamma_i$.
To estimate expectation values of all ${2N \choose 2k}$ $2k$-Majorana operators to error $\epsilon$, we need for each operator $M_i$ measurements in a basis containing this operator.
As we have established that our measurement scheme only allows such measurements in parallel if the operators commute, the bound derived in App.~\ref{app:fermionic_limits} directly bounds the number of operators that may be estimated per preparation of $\rho$ to ${N \choose k}$, and the result follows by Eq.~\ref{eq:krdm_bound}.

We now show that our measurement protocol allows only for estimation of commuting Pauli operators.
By definition, Clifford operators map Pauli operators to Pauli operators, so any measurement of a state $\rho$ that consists of a Clifford circuit $U_{\mathrm{Cl}}$ and subsequent readout in the computational basis is equivalent to a measurement of the commuting Pauli operators $\{U_{\mathrm{Cl}}^{\dag}Z_jU_{\mathrm{Cl}}\}$.
(The same is true of any tensor products $\{U_{\mathrm{Cl}}^{\dag}\otimes_jZ_jU_{\mathrm{Cl}}\}=\{\prod_jU_{\mathrm{Cl}}^{\dag}Z_jU_{\mathrm{Cl}}\}$ on $\rho$ --- where the $\otimes_j$ is taken over any set of qubits --- and the following arguments remain true if $Z_j$ is replaced by $\otimes_jZ_j$).
It remains to show that the number of preparations is unaffected by the addition of $N_a$ ancilla qubits in the $|0\rangle$ state.
Under such an addition, we may still invert the measurement $U_{\mathrm{Cl}}^{\dag}Z_jU_{\mathrm{Cl}}=P_{j,\rho}\otimes P_{j,a}$, where $P_{j,\mathrm{rho}}$ and $P_{j,a}$ are Pauli operators on the system and the ancilla qubits respectively. 
By construction, the state is separable across the bipartition into system and ancilla qubits, so $\langle P_{j,\rho}\otimes P_{j,a}\rangle = \langle P_{j,\rho}\rangle\langle P_{j,a}\rangle$.
Then, as we require our ancilla qubits to be prepared in the $|0\rangle$ state, $\langle P_{j,a}\rangle = 0$ unless $P_{j,a}$ is a tensor product of $I$ and $Z$, in which case $\langle P_{j,a}\rangle = 1$.
If $\langle P_{j,a}\rangle = 0$, a measurement of $Z_j$ does not yield any information about $\langle P_{j,\rho}\rangle$, while if $\langle P_{j,a}\rangle=1$, a measurement of $Z_j$ yields exactly the same information as a direct measurement of $P_{j,\rho}$.
Then, consider two operators $U_{\mathrm{Cl}}^{\dag}Z_jU_{\mathrm{Cl}}=P_{j,a}\otimes P_{j,\rho}$ and $U_{\mathrm{Cl}}^{\dag}Z_kU_{\mathrm{Cl}}=P_{k,a}\otimes P_{k,\rho}$.
We have that $[P_{j,a}\otimes P_{j,\rho},P_{k,a}\otimes P_{k,\rho}]$ commute, and if $\langle P_{j,a}\rangle=1$ and $\langle P_{k,a}\rangle=1$, $P_{j,a}$ and $P_{k,a}$ commute on a term-wise basis (as they are tensor products if $I$ and $Z$), which implies $[P_{j,\rho},P_{k,\rho}]=0$.
This shows that the addition of ancilla qubits in the $|0\rangle$ state cannot be used to simultaneously measure non-commuting Pauli operators via Clifford circuits, and our allowed measurements correspond to simultaneous measurement of a set of commuting Pauli operators on $\rho$, as required.

Finally, we argue for the existence of a state for which Eq.~\ref{eq:var_bound} is tight.
This may not always be the case - by constraining a fermionic $k$-RDM to the positive cone of $N$-representable states, Pauli operators with expectation values close to $\pm 1$ (and thus small variance) constrain the expectation values of anti-commuting operators near $0$ below this limit.
This beneficial covariance is of particular importance when taking linear combinations of RDM elements e.g. to calculate energies~\cite{huggins2019efficient}, however it requires a state have highly non-regular structure which in general will not be the case (nor known a priori).
The simplest example of an unstructured state is the maximally-mixed state on $N$ fermions; by definition all measurements of this state are uncorrelated, and the variance on estimation of all terms is $\mathrm{Var}(\langle\Gamma_i\rangle)=\frac{1}{4M_i}$, which achieves the upper bound in Eq.~\ref{eq:var_bound}.\qed
\end{document}